\documentclass[review,letterpaper,11pt,times]{elsarticle}

\usepackage{lineno,hyperref}
\usepackage{amsmath}
\usepackage[inline]{trackchanges}
\addeditor{EC}
\addeditor{SA}

\journal{Tectonophysics}

\biboptions{authoryear}

\begin{document}

\begin{frontmatter}

\title{Incorporating Deformation Energetics in Long-Term Tectonic Modeling}

\author[ceriaddress]{Sabber Ahamed\corref{correspondingauthor}}
\cortext[correspondingauthor]{Corresponding author}
\ead{msahamed@memphis.edu}

\author[ceriaddress]{Eunseo Choi}

\address[ceriaddress]{Center for Earthquake Research and Information, The University of Memphis, 3890 Central Ave., Memphis, TN 38152, U.S.A.}

\begin{abstract}
The deformation-related energy budget is usually considered in the simplest form or even completely omitted from the energy balance equation. We derive a full energy balance equation that accounts not only for heat energy but also for mechanical (elastic, plastic and viscous) work. The derived equation is implemented in DES3D, an unstructured finite element solver for long-term tectonic deformation. We verify the implementation by comparing numerical solutions to the corresponding semi-analytic solutions in three benchmarks extended from the classical oedometer test. Two of the benchmarks are designed to evaluate the temperature change in a Mohr-Coulomb elasto-plastic square governed by a simplified equation involving plastic power only and by the full temperature evolution equation, respectively. The third benchmark differs in that it computes thermal stresses associated with a prescribed uniform temperature increase. All the solutions from DES3D show relative errors less than 0.1 \%. We also investigate the long-term effects of deformation energetics on the evolution of large offset normal faults.
We find that the models considering the full energy balance equation tend to produce more secondary faults and an elongated core complex. Our results for the normal fault system confirm that persistent inelastic deformation has significant impact on the long-term evolution of faults, motivating further exploration of the role of the full energy balance equation in other geodynamic systems.
\end{abstract}

\begin{keyword}
plastic power\sep strain-softening plasticity\sep thermodynamic principles\sep thermal stress
\end{keyword}

\end{frontmatter}

% \linenumbers

\section{Introduction}
The energy conservation principle can describe a wider range of geological and geodynamic
phenomena when it takes into account energies involved with than the usual form involving only heat advection and diffusion
because the general form can account for the energetics of deformation, which is often important in complicated phenomena.
For instance, \citet{Hunt1991} show that
%materials softening in proportion to strain rate have a non-convex Helmholtz energy with strain as an independent variable.
non-periodic localized folding can be viewed as a superposition of folding modes
that correspond to multiple local minima of the non-convex energy function and the non-convexity originates
from deformational contribution to the system's energy budget.
On a similar note, \citet{hobbs2011thermodynamics} propose that the feedback between shear heating and
temperature-dependent viscosity can explain non-periodic and non-symmetric folding occurring in layers with small viscosity contrast,
The classical Biot's theory~\citep[e.g.,][]{Biot1961} predicts that folding would not occur in such a configuration.
Influences of energy dissipation have been considered in the lithospheric scale as well.
\citet{regenauer2006effect} consider the feedback between the energy dissipated due to inelastic deformation
and changes in viscous and plastic material properties due to temperature changes resulting from heat converted from the dissipated energy.
They find that the two-way feedback process can make the brittle-ductile transition (BDT) zone weaker than other parts of lithosphere
although the classical strength envelopes predict the BDT zone of lithosphere to be
the strongest~\citep{ranalli1987rheological,goetze1979stress,brace1980limits}.
Energy dissipated in the form of shear heating is shown to promote a necking in the subducting slab and ultimately
lead to slab detachment \citep{gerya2004thermomechanical}. In the whole-mantle scale, \citet{Yuen1987} point out that feedback between
rheology and dissipative energy in mantle convection is potentially an important mechanism that can warm up the mantle by several hundred
degrees above the incompressible profile.
% \citet{ita1994sensitivity} show that dissipative heating in the energy balance along temperature, pressure and phase dependent thermodynamic properties such as thermal expansion coefficient, heat capacity, and latent heat can significantly facilitate vertical flow across the 660-km phase boundary.%
\citet{ita1994sensitivity} show that dissipative heating in the energy balance can significantly facilitate the vertical flow across the 660-km phase boundary. %when they consider temperature, pressure, and phase dependent thermodynamic properties such as thermal expansion coefficient, heat capacity, and latent heat.

%In spite of these advances  %\add[EC]{[e.g., REFS]},
%
However, computational long-term tectonic modeling, concerned about the long-term evolution of geological structures of various scales,
has yet to fully embrace deformation energetics. Problems are centered around the fact that simplifications commonly made in this type of modeling
preclude consistent thermo-mechanical coupling. For instance, energy balance is considered only in terms of heat advection and diffusion while
thermo-mechanical feedback is realized only through temperature-dependent viscosity or shear heating.
%Even in
%In addition, const such as thermoelasticity and volumetric strain, both elastic and inelastic. However, these constitutive relations have been mostly deemed %unnecessary without justifications based on a consistent thermodynamicthermo-mechanical coupling.
%This status quo motivates us to look for ways of reconciling the differences between the prevailing conventions in LTM and the general thermodynamic frameworks.

More specifically, we identify three elements in kinematics and constitutive models that need to be incorporated into a thermodyanmic framework for long-term tectonic models.
Firstly, we note that the elastic or plastic deformations are frequently assumed to be incompressible~\citep[e.g.,][]{regenauer2003review,regenauer2006effect,regenauer2008strain,connolly2009geodynamic,hobbs2011thermodynamics} although this assumption is neither required nor well-justified.
Volumetric strain can have a significant effect on energy budget~\citep{hunsche1991volume,zinoviev1994energy} and is non-negligible during brittle deformations~\citep[e.g.][]{brace1966dilatancy} and phase transformations~\cite{hyndman2003serpentinization,hetenyi2011incorporating}.
Secondly, thermal stresses are often ignored even though they can be a significant source of transient stresses and associated deformation~\citep{choi2008thermomechanics,ChoiGurn2008,Korenaga2007}. A simple back-of-the-envelope calculation shows that a temperature change of 100 K in a perfectly confined rock body with a bulk modulus of 30 GPa can generate up to 90 MPa thermal stresses when the thermal expansion coefficient is  $3\times10^{-5} K^{-1}$. Although transient, thermal stresses of such a magnitude might be sufficient for driving permanent changes in state variables such as elastic damage or plastic strain under non-linear rheologies. %, which can affect the behavior of rocks or lithosphere in the long term.
Thirdly, strain weakening plasticity is sometimes considered as inconsistent with thermodynamic principles~\citep[e.g.][]{regenauer2006effect}. However, frictional materials do show reduction in overall strength with continued loading~\citep[e.g.][]{Read1984,Borja2000}, making it necessary to consider the softening behavior in tectonic models concerned about brittle behaviors of rocks. In fact, the strain softening is perfectly legitimate in the light of the Clausius-Duhem inequality, a statement of the 2nd law of thermodynamics~\citep[e.g., Sec. 3.2 in][]{Lubliner2008}. Rates and amounts of strain softening in rocks are still to be better constrained but reducing the uncertainty associated with them is an independent issue.

In this paper, we first derive a set of governing equations for thermo-mechanically coupled tectonic systems. We start from the generic thermodynamic principles closely following the procedure of ~\citep{wright2002physics} to derive the governing equation that incorporate the three elements discussed above. We focus on the elasto-visco-plastic material that has compressible elasticity, thermal stress and strain wakening plasticity. Integration of such kinematic and constitutive models into the general energy balance will be crucial for realistically modeling geological systems.  We then implement the derived governing equations in DES3D, an open source finite element code for geodynamic modeling, and verify the implementation semi-analytically. Finally, we explore the effects of thermo-mechanical coupling on the long-term evolution of large-offset normal faults with emphasis on the role of volumetric inelastic strain. Since extensibly studied and well understood, the normal fault systems would allow us to isolate the new effects introduced by the coupled physics.

% --------------------------------- theory ------------------------------
\section{Derivation of governing equations}

\subsection{Energy Balance Equation}

Several theoretical works have derived a set of governing equations for a thermo-mechanical system from the general form of the thermodynamic principles. The common procedure is to relate the rate of change of the internal energy appearing in the statement of energy balance to that of thermodynamic potentials such as the Helmholtz free energy and the Gibbs free energy. Thermodynamic potentials involve the capacity to do mechanical work in addition to heat content. For instance, the Helmholtz free energy is ``the portion of the internal energy available for doing work at constant temperature''~\citep[p.263 in][]{Malvern1969}. The main difference between the the Helmholtz and the Gibbs free energy is whether strain is an independent variable as in the former or stress is as in the latter. Since the definitions of these thermodynamic potentials involve the product of temperature and entropy, the energy balance principle takes an intermediate form involving the time derivative of entropy. The last step in deriving the temperature evolution equation is to express the time derivative of entropy in terms of that of temperature and other variables. \citet{Regenauer_Lieb_2003} start from the energy conservation principle stated in terms of the Helmholtz free energy to derive the partial differential equation for temperature evolution as well as other equations that are coupled with it (e.g., mass conservation equation and constitutive relations) for shear zone-developing systems in the Earth and other planets. Their final system of equations can consistently describe the feedback processes among energy, rheology and other variables such as grain size and water content in shear zone formation. Similarly, \citet{lyakhovsky1997distributed} show how an evolution equation for elastic damage can be derived from the energy conservation principle. Also starting from the energy balance equation in terms of the Helmholtz free energy, they include a non-dimensional variable quantifying the amount of damage along with temperature and infinitesimal elastic strain as independent variables of the free energy. By treating damage process as a source of entropy in the intermediate equation for entropy evolution, they derive a damage evolution equation that is proportional to the rate of free energy change with damage.

For completeness, we derive a temperature evolution equation from the generic energy balance principle involving the Gibbs free energy ($g$), following ~\citet{wright2002physics}.
This form of the energy balance principle is different from those of related studies in geodynamics ~\citep{regenauer2003review, lyakhovsky1997distributed} in that those earlier studies used another thermodynamic potential, the Helmholtz free energy.
In the context of continuum thermodynamics, the Gibbs free energy
is a function of Cauchy stress ($\boldsymbol{\sigma}$), temperature ($T$)
and a set of internal variables ($q_j$, $j=0,1,2...n$) while the Helmholtz energy has elastic strain
in place of Cauchy stress. The work by~\citet{wright2002physics} inspired our choice of thermodynamic
potential %, banerjee2007basic}.
but one difference is that we assume infinitesimal strain while~\citet{wright2002physics} used the finite strain kinematics.

%%
%%--------------------- equation ------------------------------
%\begin{equation} \label{1}
%g = g(\boldsymbol{\sigma}, \, T, \, q_j),
%\end{equation}
%%--------------------------------------------------------------
%
%Applying our small deformation assumptions
Additive strain rate decomposition is allowed under our assumption of infinitesimal strain. Furthermore, the class of material we are interested in is elasto-visco-plastic. Under these assumptions, total strain rate is decomposed as follows:
%
%--------------------- equation ------------------------------
\begin{equation} \label{1.5}
\dot{\boldsymbol{\epsilon}} = \dot{\boldsymbol{\epsilon}}_e + \dot{\boldsymbol{\epsilon}}_p +  \dot{\boldsymbol{\epsilon}}_v,
\end{equation}
where $\dot{\boldsymbol{\epsilon}}_e$, $\dot{\boldsymbol{\epsilon}}_p$, and $\dot{\boldsymbol{\epsilon}}_v$ are elastic, plastic and viscous strain rate, respectively.

Following~\citet{wright2002physics}, we define the Gibbs free energy per unit mass as
%
%--------------------- equation ------------------------------
\begin{equation} \label{2}
%g(\boldsymbol{\sigma}, \, T, \, q_j) = -e +T\eta +\frac{1}{\q_0} \boldsymbol{\sigma}:\boldsymbol{\epsilon}_e,
g(\boldsymbol{\sigma}, \, T, \, q_j) = -e +T\eta + \frac{1}{\rho} \, \boldsymbol{\sigma}:\boldsymbol{\epsilon}_e,
\end{equation}
%--------------------------------------------------------------
%
where $e$ is the internal energy per unit mass, $\eta$ is entropy per unit mass, $\rho$ is density and $\boldsymbol{\epsilon}_e$ is elastic strain tensor. Only the elastic strain appears in the definition because it is the only component of strain that can contribute to stored energy. Taking the totral differential of $g$ and rearranging terms, we get the differential of internal energy:%Differentiating the equation (\ref{2}) gives
%
%--------------------- equation ------------------------------
\begin{equation} \label{4}
%de = -dg + \eta \, dT + Td\eta -\frac{1}{q^2_0}\boldsymbol{\sigma}:\boldsymbol{\epsilon}_e dq_0 + \frac{1}{q_0}\boldsymbol{\epsilon}_e: d\boldsymbol{\sigma} +\frac{1}{q_0}\boldsymbol{\sigma}:d\boldsymbol{\epsilon}_e.
de = -dg + \eta \, dT + Td\eta - \frac{1}{\rho^{2}} \, \boldsymbol{\sigma}:\boldsymbol{\epsilon}_{e} d\rho + \frac{1}{\rho} \, \boldsymbol{\epsilon}_e: d\boldsymbol{\sigma} + \frac{1}{\rho} \, \boldsymbol{\sigma}:d\boldsymbol{\epsilon}_e.
\end{equation}
%--------------------------------------------------------------
%
Since $g$ = $g(\boldsymbol{\sigma}, T, q_{j})$, the total differential of $g$ is also
%
%--------------------- equation ------------------------------
\begin{equation} \label{5}
dg = \frac{\partial g}{\partial \boldsymbol{\sigma}}:d\boldsymbol{\sigma}+ \frac{\partial  g}{\partial T} \, dT
+
\sum\limits_{j=0}^n \frac{\partial  g}{\partial q_{j}} d q_{j},
\end{equation}
where $q_{0} = \rho$ according to the conventions of~\citet{wright2002physics}.
%--------------------------------------------------------------
%
Elastic strain and entropy \citep[c.f.,][]{wright2002physics} are defined as
%
%--------------------- equation ------------------------------
\begin{equation} \label{6}
%\boldsymbol{\epsilon}_e  =  q_0 \frac{\partial g}{\partial \boldsymbol{\sigma}}\enspace \text{and}\enspace \eta =\frac{\partial g}{\partial T }.
\boldsymbol{\epsilon}_e  \equiv  \rho \frac{\partial g}{\partial \boldsymbol{\sigma}}\enspace \text{and} \enspace
\eta \equiv \frac{\partial g}{\partial T }.
\end{equation}
%--------------------------------------------------------------
%
From the equations (\ref{4}), (\ref{5}) and (\ref{6}) we get
%
%--------------------- equation ------------------------------
\begin{equation} \label{8}
de= -\sum\limits_{j=0}^n\frac{\partial g}{\partial q_j}dq_j+ Td\eta-\frac{1}{\rho^{2}}\boldsymbol{\sigma}:\boldsymbol{\epsilon}_e d\rho + \frac{1}{\rho}\boldsymbol{\sigma}: d\boldsymbol{\epsilon}_e
\end{equation}
%--------------------------------------------------------------
%
Since $q_{0}=\rho$, the terms containing $dq_0$ and $d\rho$ can be grouped together. With the grouping, equation (\ref{8}) becomes
%
%--------------------- equation ------------------------------
\begin{equation}\nonumber
de= \frac{1}{\rho} \, \boldsymbol{\sigma}:d\boldsymbol{\epsilon}_e+ Td\eta- \left( \frac{\partial g}{\partial \rho}+\frac{1}{\rho^{2}} \boldsymbol{\sigma}:\boldsymbol{\epsilon}_e \right) d\rho - \sum\limits_{j=1}^n \frac{\partial g}{\partial q_j} \, dq_j.
\end{equation}
%--------------------------------------------------------------
%
With two new notations
%
%--------------------- equation ------------------------------
\begin{equation} \label{8.1}
Q_0 \equiv - \left( \rho \frac{\partial g}{\partial \rho }+\frac{1}{\rho}\boldsymbol{\sigma}: \boldsymbol{\epsilon}_e \right)
\end{equation}
%--------------------------------------------------------------
%
and
%--------------------- equation ------------------------------
\begin{equation} \label{8.2}
Q_j \equiv -\rho \, \frac{\partial g}{\partial q_j} \quad \text{(j=1, 2, 3, $\ldots$)},
\end{equation}
equation (\ref{8}) is simplified to
%--------------------- equation ------------------------------
\begin{equation} \label{9}
de= \frac{1}{\rho}\boldsymbol{\sigma}:d\boldsymbol{\epsilon}_e+ Td\eta + \frac{1}{\rho}\sum\limits_{j=0}^n Q_j \, dq_j.
\end{equation}
%
% The equation (\ref{9}) shows that the internal energy ($e$) is a function of
% %
% %--------------------- equation ------------------------------
% \begin{equation} \label{10}
% e= e(\boldsymbol{\epsilon}_e, \eta, q_0, q_j)
% \end{equation}
%
Differentiating the equation (\ref{9}) with respect to time ($t$) we have
%
%--------------------- equation ------------------------------
\begin{equation} \label{11}
\frac{de}{dt}
=
T \, \frac{d\eta}{dt} + \frac{1}{\rho} \, \boldsymbol{\sigma}:\dot{\boldsymbol{\epsilon}}_e
+
\frac{1}{\rho}\sum\limits_{j=0}^n Q_j \frac{d q_j}{dt}.
\end{equation}
Multiplying the equation (\ref{11}) by $\rho$, we get the following equation for the time rate of change of internal energy per volume:
%
%--------------------- equation ------------------------------
\begin{equation} \label{12}
\rho\frac{d e}{d t}
=
\rho \, T \, \frac{d\eta}{dt}
+
\boldsymbol{\sigma}:\dot{\boldsymbol{\epsilon}}_e + \sum\limits_{j=0}^n Q_j \frac{d q_j}{dt}.
\end{equation}
%
%We have derived the time derivative of the internal energy.

To relate the material time derivative of the internal energy given in~\eqref{12} to the energy balance principle, we recall the general form of the energy balance equation~\citep[e.g.,][]{kennett2008geophysical, Malvern1969, wright2002physics}:
%
%--------------------- equation ------------------------------
\begin{equation} \label{13}
\rho \frac{de}{dt} =\boldsymbol{\sigma}:\nabla \boldsymbol{v} - \nabla \cdot \boldsymbol{q} + \rho \, s,
\end{equation}
where $s$ is a heat energy source or sink per mass, $\boldsymbol{q}$ is heat flux and $\boldsymbol{v}$ is velocity.
According to the additive decomposition of strain rate in~\eqref{1.5},
%--------------------- equation ------------------------------
\begin{equation} \label{14}
\boldsymbol{\sigma}:\dot{\boldsymbol{\epsilon}}_e = \boldsymbol{\sigma}:\dot{\boldsymbol{\epsilon}} - \boldsymbol{\sigma}:(\dot{\boldsymbol{\epsilon}}_p + \dot{\boldsymbol{\epsilon}}_v).
\end{equation}
Since the double dot product of the symmetric $\boldsymbol{\sigma}$ and the anti-symmetric part of $\nabla \boldsymbol{v}$ is zero, $\boldsymbol{\sigma}:\nabla \boldsymbol{v} = \boldsymbol{\sigma}:\dot{\boldsymbol{\epsilon}}$. By eliminating the time derivative of internal energy from equations (\ref{12}) and (\ref{13}) and then using (\ref{14}), we get
%
%--------------------- equation ------------------------------
\begin{equation}\nonumber
\rho \, T \, \frac{d \eta}{dt}
+
\boldsymbol{\sigma}:\dot{\boldsymbol{\epsilon}}
-
\boldsymbol{\sigma}:(\dot{\boldsymbol{\epsilon}}_p
+
\dot{\boldsymbol{\epsilon}}_v)
+
\sum\limits_{j=0}^nQ_j  \frac{d q_j}{dt}
-
\boldsymbol{\sigma}:\dot{\boldsymbol{\epsilon}} - \rho \, s + \nabla \cdot \boldsymbol{q} = 0,
\end{equation}
which is simplified to
%--------------------- equation ------------------------------
\begin{equation} \label{15}
\rho \, T \, \frac{d\eta}{dt}
=
\boldsymbol{\sigma}:(\dot{\boldsymbol{\epsilon}}_p
+
\dot{\boldsymbol{\epsilon}}_v)
-
\nabla \cdot \boldsymbol{q}
+ \rho \, s
-
\sum\limits_{j=0}^nQ_j\frac{d q_j}{dt}.
\end{equation}

The final step to derive a partial differential equation for temperature from~\eqref{15} is to relate the time derivative of entropy per mass to temperature. The equipresence principle ~\citep{Malvern1969} requires the entropy $\eta$ to have the same set of independent variables as the Gibbs free energy. In other words, the entropy per unit mass is also a function of the Cauchy stress ($\boldsymbol{\sigma}$), the temperature ($T$) and a set of internal variables ($q_j$, $j=0,1,2 ... n$). The total differential of $\eta =\eta (\boldsymbol{\sigma}, T, q_{j})$ is
%
%--------------------- equation ------------------------------
\begin{equation}\nonumber
d\eta =\frac{\partial \eta}{\partial \boldsymbol{\sigma}}: d\boldsymbol{\sigma} +\frac{\partial \eta}{\partial T} \, dT +\sum\limits_{j=0}^n \frac{\partial \eta}{\partial q_j} \, dq_j.
\end{equation}
Identifying the first internal variable ($q_{0}$) with density ($\rho$) again, we get
%--------------------- equation ------------------------------
\begin{equation} \label{16}
d\eta =\frac{\partial \eta}{\partial \boldsymbol{\sigma}}: d\boldsymbol{\sigma} +\frac{\partial \eta}{\partial T} dT+\frac{\partial \eta}{\partial \rho} d\rho+\sum\limits_{j = 1}^n \frac{\partial \eta}{\partial q_j} dq_j.
\end{equation}
When differentiated with respect to time, the equation (\ref{16}) becomes
%
%--------------------- equation ------------------------------
\begin{equation} \label{17}
\frac{d\eta}{dt}  =\frac{\partial \eta}{\partial \boldsymbol{\sigma}}:\frac{d \boldsymbol{\sigma}}{dt} +\frac{\partial \eta}{\partial T} \frac{ dT}{dt}+\frac{\partial \eta}{\partial \rho} \frac{d \rho}{dt}+\sum\limits_{j=1}^n \frac{\partial \eta}{\partial q_j} \frac{d q_j}{dt}.
\end{equation}
According to~\eqref{6},
%
%--------------------- equation ------------------------------
\begin{equation} \label{17.1}
\frac{\partial \eta}{\partial \boldsymbol{\sigma}} = \frac{\partial}{\partial \boldsymbol{\sigma}} \frac{\partial g}{\partial T}
= \frac{\partial}{\partial T} \frac{\partial g}{\partial \boldsymbol{\sigma}} = \frac{1}{\rho}\frac{\partial \boldsymbol{\epsilon}_e}{\partial T}.
\end{equation}
From the definition of the specific heat at constant stress, $c_{\boldsymbol{\sigma}}$, we get the following identity:
%
%--------------------- equation ------------------------------
\begin{equation} \label{17.4}
 c_{\boldsymbol{\sigma}} \equiv \frac{\partial e}{\partial T} = \frac{\partial}{\partial T} \left( -g + T\,\eta + \frac{1}{\rho} \boldsymbol{\sigma}:\boldsymbol{\epsilon}_{e} \right) = T \frac{\partial \eta}{\partial T} + \frac{1}{\rho} \boldsymbol{\sigma}:\frac{\partial \boldsymbol{\epsilon}_{e}}{\partial T}.
\end{equation}
Using~\eqref{17.4}, we get
%
%--------------------- equation ------------------------------
\begin{equation} \label{17.5}
 \frac{\partial \eta}{\partial T} = \frac{1}{T}\left(c_{\boldsymbol{\sigma}}- \frac{1}{\rho}\boldsymbol{\sigma}:\frac{\partial \boldsymbol{\epsilon}_e}{\partial T}\right).
\end{equation}
For convenience, we express partial derivatives of entropy per mass with respect to density and other internal variables in terms of $Q_{0}$ and $Q_{j}$ ($j = 1, 2, 3, \ldots$), which are defined in \eqref{8.1} and \eqref{8.2}.
%
%--------------------- equation ------------------------------
\begin{equation} \label{17.2}
\frac{\partial \eta}{\partial \rho} = \frac{\partial}{\partial \rho} \frac{\partial g}{\partial T}
= \frac{\partial}{\partial T} \frac{\partial g}{\partial \rho} = -\frac{1}{\rho}\frac{\partial Q_0}{\partial T}-\frac{1}{\rho^2}\boldsymbol{\sigma}:\frac{\partial \boldsymbol{\epsilon}_e}{\partial T}.
\end{equation}
and
%
%--------------------- equation ------------------------------
\begin{equation} \label{17.3}
\frac{\partial \eta}{\partial q_{j}} = \frac{\partial}{\partial q_{j}} \frac{\partial g}{\partial T}
= \frac{\partial}{\partial T} \frac{\partial g}{\partial q_{j}} = -\frac{1}{\rho} \frac{\partial Q_j}{\partial T},
\end{equation}
for $j = 1, 2, 3, \ldots$.
%
%
%Since From~\eqref{5} and~\eqref{6}, the differential of $g$ is given by
%--------------------- equation ------------------------------
%\begin{equation} \label{17.6}
%  dg = \frac{1}{\rho} \boldsymbol{\epsilon}_{e} : d\boldsymbol{\sigma} + \eta \, c_{\boldsymbol{\sigma}} = \frac{\partial e}{\partial T} = \frac{\partial}{\partial T} \left( -g + T\,\eta + \frac{1}{\rho} \boldsymbol{\sigma}:\boldsymbol{\epsilon}_{e} \right) = T \frac{\partial \eta}{\partial T} + \frac{1}{\rho} \boldsymbol{\sigma}:\frac{\partial \boldsymbol{\epsilon}_{e}}{\partial T}.
%\end{equation}
%
%From the entropy potential~\citep[p. 17 in][]{wright2002physics} we have
%
%--------------------- equation ------------------------------
%\begin{equation} \label{18}
%\begin{split}
%\frac{\partial \eta}{\partial \boldsymbol{\sigma}} &= \frac{\partial}{\partial \boldsymbol{\sigma}} \frac{\partial g}{\partial T} = \frac{1}{\rho}\frac{\partial \boldsymbol{\epsilon}_e}{\partial T}, \\
%\text{; }
%\frac{\partial \eta}{\partial T} &= \frac{1}{T}\left(c_{\boldsymbol{\sigma}}- \frac{1}{\rho}\boldsymbol{\sigma}:\frac{\partial \boldsymbol{\epsilon}_e}{\partial T}\right),\\
%
%\frac{\partial \eta}{\partial \rho} &= -\frac{1}{\rho}\frac{\partial Q_0}{\partial T}-\frac{1}{\rho^2}\boldsymbol{\sigma}:\frac{\partial \boldsymbol{\epsilon}_e}{\partial T},\\
%
%\text{and}\enspace \frac{\partial \eta}{\partial q_j} &= -\frac{1}{\rho} \frac{\partial Q_j}{\partial T}, \\
%
%\end{split}
%\end{equation}
%
%where $c_p$ is the specific heat at constant pressure.
Plugging \eqref{17.1}, \eqref{17.5}, \eqref{17.2} and \eqref{17.3} and into the equation \eqref{17}, we get
%
%--------------------- equation ------------------------------
\begin{equation}\nonumber
\begin{split}
\frac{d \eta}{dt} = &
\frac{1}{\rho}\frac{\partial \boldsymbol{\epsilon}_e}{\partial T} : \frac{d\boldsymbol{\sigma}}{dt}
+
\frac{1}{T}\left(c_{\boldsymbol{\sigma}} - \frac{1}{\rho}\boldsymbol{\sigma}:\frac{\partial \boldsymbol{\epsilon}_e}{\partial T}\right)\frac{dT}{dt} \\
&+
\left( -\frac{1}{\rho} \frac{\partial Q_0}{\partial T} - \frac{1}{\rho^2}\boldsymbol{\sigma}:\frac{\partial \boldsymbol{\epsilon}_e}{\partial T} \right) \frac{d\rho}{dt}
-
\frac{1}{\rho}\sum\limits_{j=1}^n\frac{\partial Q_j}{\partial T}\frac{d q_j}{dt}.
\end{split}
\end{equation}
This equation can be further simplified to
%--------------------- equation ------------------------------
\begin{equation} \label{19}
\begin{split}
\frac{d\eta}{dt}
=
\frac{1}{T} \left( c_{\boldsymbol{\sigma}} - \frac{1}{\rho}\boldsymbol{\sigma}:\frac{\partial \boldsymbol{\epsilon}_e}{\partial T} \right) \frac{dT}{dt}
+
\frac{1}{\rho}\frac{\partial \boldsymbol{\epsilon}_e}{\partial T}: \frac{d \boldsymbol{\sigma}}{dt}
& -
\frac{1}{\rho^2} \left( \boldsymbol{\sigma}:\frac{\partial \boldsymbol{\epsilon}_e}{\partial T} \right) \frac{d\rho}{dt} \\
& -
\frac{1}{\rho}\sum\limits_{j=0}^n\frac{\partial Q_j}{\partial T}\frac{d q_j}{dt}.
\end{split}
\end{equation}
Substituting the equation (\ref{19}) into the equation (\ref{15}) we get:
%
%--------------------- equation ------------------------------
\begin{equation}\nonumber
\begin{split}
\rho T\left[\frac{1}{T}\left(c_{\boldsymbol{\sigma}}- \frac{1}{\rho}\boldsymbol{\sigma}:\frac{\partial \boldsymbol{\epsilon}_e}{\partial T}\right)\frac{dT}{dt}
+
\frac{1}{\rho}\frac{\partial \boldsymbol{\epsilon}_e}{\partial T}:\left(\frac{d\boldsymbol{\sigma}}{dt}-\frac{1}{\rho}\boldsymbol{\sigma}\frac{d \rho}{dt}\right)
-
 \frac{1}{\rho}\sum\limits_{j=0}^n\frac{\partial Q_j}{\partial T}\frac{d q_j}{dt}\right]\\
 -
 \boldsymbol{\sigma}:(\dot{\boldsymbol{\epsilon}}_p + \dot{\boldsymbol{\epsilon}}_v) + \sum\limits_{j=0}^nQ_j\frac{d q_j}{dt}-\rho s
 +
  \nabla \cdot \boldsymbol{q} = 0,
\end{split}
\end{equation}
%
%--------------------- equation ------------------------------
% \begin{equation}\nonumber
% \begin{split}
% \rho\left(c_p- \frac{1}{\rho}\boldsymbol{\sigma}:\frac{\partial \boldsymbol{\epsilon}_e}{\partial T}\right)\frac{\partial T}{\partial t} +\frac{\rho}{\rho }T \frac{\partial \boldsymbol{\epsilon}_e}{\partial T}:\left(\frac{\partial \boldsymbol{\sigma}}{\partial t}-\frac{1}{\rho}\boldsymbol{\sigma}\frac{\partial q_j}{\partial t}\right)-T\sum\limits_{j=1}^n\frac{\partial Q_j}{\partial T}\frac{\partial q_j}{\partial t}\\+ \sum\limits_{j=1}^n Q_j\frac{\partial q_j}{\partial t}-\boldsymbol{\sigma}:(\dot{\boldsymbol{\epsilon}}_p + \dot{\boldsymbol{\epsilon}}_v) -\rho s + \nabla.q=0,
% \end{split}
% \end{equation}
%
which is simplified to
%--------------------- equation ------------------------------
\begin{equation} \label{20}
\begin{split}
& \rho\left(c_{\boldsymbol{\sigma}} - \frac{1}{\rho}\boldsymbol{\sigma}:\frac{\partial \boldsymbol{\epsilon}_e}{\partial T}\right)\frac{dT}{dt}
+
T \frac{\partial \boldsymbol{\epsilon}_e}{\partial T}:\left(\frac{d\boldsymbol{\sigma}}{dt}-\frac{1}{\rho}\boldsymbol{\sigma}\frac{d\rho}{dt}\right) \\
&+
\sum\limits_{j=0}^n\left(Q_j-T\frac{\partial Q_j}{\partial T}\right)\frac{d q_j}{dt}
-
\boldsymbol{\sigma}:(\dot{\boldsymbol{\epsilon}}_p + \dot{\boldsymbol{\epsilon}}_v) -\rho s + \nabla \cdot \boldsymbol{q} = 0.
\end{split}
\end{equation}

Strain due to temperature change is assumed to be isotropic: i.e., $\partial \boldsymbol{\epsilon}_e / \partial T=\alpha_{l} \boldsymbol{I}$, where $\boldsymbol{I}$ is the identity tensor and $\alpha_{l}$ is linear thermal expansion coefficient and 1/3 of the volumetric thermal expansion coefficient, $\alpha_{v}$. Under this assumption, the equation (\ref{20}) is rearranged as follows:
%
%--------------------- equation ------------------------------
% \begin{equation}
% \begin{split}
% \rho\left(c_p- \frac{1}{\rho}\boldsymbol{\sigma}:\alpha \boldsymbol{I}\right)\frac{\partial T}{\partial t} = -\nabla \cdot q + \rho s +\boldsymbol{\sigma}:(\dot{\boldsymbol{\epsilon}}_p + \dot{\boldsymbol{\epsilon}}_v) -T\alpha \boldsymbol{I}:\left(\frac{\partial \boldsymbol{\sigma}}{\partial t}-\frac{1}{\rho}\boldsymbol{\sigma}\frac{\partial \rho}{\partial t}\right)\\-\sum\limits_{j=0}^n\left(Q_j-T\frac{\partial Q_j}{\partial T}\right)\frac{\partial q_j}{\partial t},
% \end{split}
% \end{equation}
%
%--------------------- equation ------------------------------
\begin{equation} \label{21}
\begin{split}
\rho c_{\boldsymbol{\sigma}} \frac{d T}{dt} - \boldsymbol{\sigma}:\alpha_{l} \boldsymbol{I}\frac{dT}{dt}= &
-
\nabla \cdot \boldsymbol{q} + \rho s
+
\boldsymbol{\sigma}:(\dot{\boldsymbol{\epsilon}}_p + \dot{\boldsymbol{\epsilon}}_v) \\
& -
T \alpha_{l} \boldsymbol{I}: \left( \frac{d \boldsymbol{\sigma}}{dt}- \frac{1}{\rho}\frac{d\rho}{dt}\boldsymbol{\sigma} \right) \\
& -
\sum\limits_{j=0}^n\left(Q_j-T\frac{\partial Q_j}{\partial T}\right)\frac{d q_j}{dt}.
\end{split}
\end{equation}
Using the identity $\boldsymbol{\sigma}:\alpha_{l} I= -3 \, \alpha_{l} \, p = -\alpha_{v} \, p$ where $p= - \frac{1}{3} \sigma_{ii}$, we have the following final form of the energy balance equation:
%
%--------------------- equation ------------------------------
\begin{equation} \label{22}
\begin{split}
(\rho c_p+ p\,\alpha_{v})\frac{dT}{dt}
= &
-\nabla \cdot \boldsymbol{q}  + \rho s +\boldsymbol{\sigma}:(\dot{\boldsymbol{\epsilon}}_p + \dot{\boldsymbol{\epsilon}}_v)
+
T \alpha_{v} \frac{dp}{dt} \\
& -
p\,T\frac{\alpha_{v}}{\rho}\frac{d \rho}{dt}
-
\sum\limits_{j=0}^n\left(Q_j-T\frac{\partial Q_j}{\partial T}\right)\frac{d q_j}{dt}.
\end{split}
\end{equation}

The last term of \eqref{22} corresponds to the changes in energy due to internal variables only.
These terms are often parametrized into a coefficient for the inelastic
power term as in $\chi \boldsymbol{\sigma}:(\dot{\boldsymbol{\epsilon}}_p + \dot{\boldsymbol{\epsilon}}_v)$
~\citep{regenauer2003review, wright2002physics}. With this parametrization, the equation \eqref{22} is simplied to
%--------------------- equation ------------------------------
\begin{equation} \label{23}
(\rho c_p+ p\,\alpha_{v})\frac{dT}{dt}
=
-\nabla \cdot \boldsymbol{q}  + \rho s +\chi \boldsymbol{\sigma}:(\dot{\boldsymbol{\epsilon}}_p + \dot{\boldsymbol{\epsilon}}_v)
+
T \alpha_{v} \frac{dp}{dt}
-
p\,T\frac{\alpha_{v}}{\rho}\frac{d \rho}{dt}.
\end{equation}
$\chi$ close to 1 means that the energy change due
to changes in internal variables is negligible relative to inelastic power and most of the inelastic power is
converted to heat production. To our knowledge, the value of $\chi$ is not very well constrained for rocks
but at least for metals, it is either close to 1 or saturates towards 1 with increasing plastic strain~\citep{wright2002physics}.
\citet{regenauer2003review} used 0.85 as the value of $\chi$. $\chi$ is assumed to be 1 in this study.
%Expanding the last term of \eqref{22} and using the definition of $Q_{0}$ given in \eqref{8.1}, we can get a slightly different form that more clearly separates internal variable terms that can be easily quantified from more obscure terms involving partial derivatives of $g$ with respect to internal variables, $q_{j}$'s:
%%
%%--------------------- equation ------------------------------
%\begin{equation} \label{23}
%\begin{split}
%(\rho c_p+ p\,\alpha_{v})\frac{dT}{dt}
%= &
%-\nabla \cdot \boldsymbol{q}  + \rho s +\boldsymbol{\sigma}:(\dot{\boldsymbol{\epsilon}}_p + \dot{\boldsymbol{\epsilon}}_v)
%+
%T \alpha_{v} \frac{dp}{dt}
%+
%( \boldsymbol{\sigma}:\boldsymbol{\epsilon}_{e}) \frac{1}{\rho} \frac{d \rho}{dt} \\
%& +
%\left( \rho \frac{\partial g}{\partial \rho} - \rho T \frac{\partial^{2} g}{\partial T \partial \rho} \right) \frac{d \rho}{dt}
%-
%\sum\limits_{j=1}^n\left(Q_j-T\frac{\partial Q_j}{\partial T}\right)\frac{d q_j}{dt}.
%\end{split}
%\end{equation}
%
%
%\note[EC]{I'll add one note on the internal variable terms.}

%***************** Solving mass conservation ******************
\subsection{Mass balance equation}
%**************************************************************

The form of the mass conservation equation involving the material derivative is
%
%--------------------- equation ------------------------------
\begin{equation} \label{24}
  \frac{d \rho}{d t}  = -\rho \nabla \cdot \boldsymbol{v}.
\end{equation}
In the Lagrangian description of motion that we are going to adopt, the above time derivative is understood as the partial derivative with respect to time, not as the material time derivative. We substitute~\eqref{24} into the last term of~\eqref{23} to get
%
%--------------------- equation ------------------------------
\begin{equation} \label{25}
  (\rho c_p+ p\,\alpha_{v})\frac{dT}{dt} = -\nabla \cdot \boldsymbol{q}  + \rho s + \boldsymbol{\sigma}:(\dot{\boldsymbol{\epsilon}}_p + \dot{\boldsymbol{\epsilon}}_v) + T \alpha_{v} \frac{dp}{dt} + p\,T \alpha_{v} \nabla \cdot \boldsymbol{v}.
\end{equation}
%
%and
%%
%%--------------------- equation ------------------------------
%\begin{equation} \label{26}
%\begin{split}
%(\rho c_p+ p\,\alpha_{v})\frac{dT}{dt}
%= &
%-\nabla \cdot \boldsymbol{q}  + \rho s +\boldsymbol{\sigma}:(\dot{\boldsymbol{\epsilon}}_p + \dot{\boldsymbol{\epsilon}}_v)
%+
%T \alpha_{v} \frac{dp}{dt}
%-
%( \boldsymbol{\sigma}:\boldsymbol{\epsilon}_{e}) \nabla \cdot \boldsymbol{v} \\
%& +
%\left( \rho \frac{\partial g}{\partial \rho} - \rho T \frac{\partial^{2} g}{\partial T \partial \rho} \right) \frac{d \rho}{dt}
%-
%\sum\limits_{j=1}^n\left(Q_j-T\frac{\partial Q_j}{\partial T}\right)\frac{d q_j}{dt}.
%\end{split}
%\end{equation}

%%***************** Solving momentum balance ******************
%\subsection{Momentum balance equation}
%%**************************************************************
%
%The form of the mass conservation equation involving the material derivative is
%%
%%--------------------- equation ------------------------------
%%\begin{equation} \label{24}
%%\frac{d \rho}{d t}  = -\rho \nabla \cdot \boldsymbol{v}.
%%\end{equation}
%%
%It is sufficient to solve the mass balance equation in a thermo-mechanically coupled system without explicitly relating the rate of density change to temperature or stress changes because the velocity divergence term in \eqref{24} will implicitly reflect such relations.

%*********** Inclusion of Thermal stress *****************
\subsection{Thermoelastic constitutive equations}
%The total elastic strains at each point of a heated body are made up of two parts. The first part comprises the strains required to maintain the continuity of the body as well as those arising because of external loads. The second part is a uniform expansion proportional to the temperature rise ($\delta T$) from some reference temperature, i.e., $\delta T = T-T_0$. If the coefficient of linear thermal expansion is detonated by $\alpha_l = \frac{1}{3}\alpha_v$ where $\alpha_v$ is volumetric coefficient of thermal expansion, then the normal thermal strain in any direction is equal to $\alpha_l \delta T$
%% ~\citep{boley2012theory, turcotte2014geodynamics}.
%
%Therefore, the total strains of a heated body is given as
%%
%%--------------------- equation ------------------------------
%\begin{align} \label{28}
%\boldsymbol{\epsilon}_{xx}&=\frac{1}{E}\left[\boldsymbol{\sigma}_{xx}-\nu(\boldsymbol{\sigma}_{yy}+\boldsymbol{\sigma}_{zz})\right] + \frac{1}{3}\alpha_v \delta T, \notag \\
%\boldsymbol{\epsilon}_{yy}&=\frac{1}{E}\left[\boldsymbol{\sigma}_{yy}-\nu(\boldsymbol{\sigma}_{zz}+\boldsymbol{\sigma}_{xx})\right] + \frac{1}{3}\alpha_v \delta T, \notag \\
%\boldsymbol{\epsilon}_{zz}&=\frac{1}{E}\left[\boldsymbol{\sigma}_{zz}-\nu(\boldsymbol{\sigma}_{xx}+\boldsymbol{\sigma}_{yy})\right] + \frac{1}{3}\alpha_v \delta T,
%\end{align}
%%
%where $\nu$ and $E$ are the Poisson's ratio and the Young's modulus respectively. It is sometimes convenient to express the stress explicitly in terms of strains
Assuming the linear isotropic elasticity and temperature change $\delta T$, we use the following thermoelastic constitutive equations~\citep[e.g.,][]{boley2012theory}:
%
%--------------------- equation ------------------------------
\begin{align} \label{29}
\sigma_{ij} = \lambda \varepsilon_{kk} \delta_{ij} + 2 G \varepsilon_{ij} - K \alpha_{v} \delta T,
%\boldsymbol{\sigma}_{xx}&= \lambda(\boldsymbol{\epsilon}_{xx}+\boldsymbol{\epsilon}_{yy}+\boldsymbol{\epsilon}_{zz})+2G\boldsymbol{\epsilon}_{xx} - K \alpha_v \delta T, \notag \\
%\boldsymbol{\sigma}_{yy}&= \lambda(\boldsymbol{\epsilon}_{xx}+\boldsymbol{\epsilon}_{yy}+\boldsymbol{\epsilon}_{zz})+2G\boldsymbol{\epsilon}_{yy} - K \alpha_v \delta T, \notag \\
%\boldsymbol{\sigma}_{zz}&= \lambda(\boldsymbol{\epsilon}_{xx}+\boldsymbol{\epsilon}_{yy}+\boldsymbol{\epsilon}_{zz})+2G\boldsymbol{\epsilon}_{zz} - K \alpha_v \delta T,
\end{align}
where $\lambda$ and $G$ are the Lam\'{e}'s constant and $K$ is the bulk modulus defined as $\lambda+2G/3$. These thermal elastic constitutive equations become the basis for viscoelastic or elastoplastic constitutive models as described in \citep{choi2013dynearthsol2d}.

\subsection{On the use of non-associated strain-softening plasticity}
%Criticism on the use of strain weakening as an inducing mechanism for strain localization is usually directed towards the imposition of \emph{ad hoc} rules of strain softening [e.g., Regenaur-Lieb et al., 2003; 2006; 2008(?)]. We agree that strain-softening rules often used tectonic modeling are lack of reliable experimental or observational basis. However, we can argue at least that the lack of constraints on the softening rate does not necessarily invalidate the shear band formation induced by strain softening, which is perfectly legitimate in the light of the 2nd law of thermodynamics. Following Belytschko et al. (2013), for instance, the Clausius-Duhem inequality boils down to the plastic dissipation inequality:
%\begin{equation*}
%  \boldsymbol{\sigma} : \boldsymbol{D}^{p} - \rho \frac{\partial \Psi}{\partial \xi_{\alpha}} \dot{\xi_{\alpha}} \ge 0.
%\end{equation*}
%
%Under some simplifying assumptions for the internal variable, which are also applied to this study, the above inequality is further simplified to [Maugin, 1992]
%\begin{equation*}
%  \boldsymbol{\sigma} : \boldsymbol{D}^{p} \ge 0.
%\end{equation*}
%
%This form of the 2nd law of thermodynamics applies to strain hardening as well as strain softening materials as nicely illustrated in Fig. 3.2.3 of [Lubliner, 2008] (see also, Sec.5.10.9, Belytschko et al., 2013].
%
Using a non-associated, strain-weakening plasticity for inducing shear bands as in~\citep{choi2013dynearthsol2d} has been described as if it violated the 2nd law of thermodynamics~\citep[e.g.,][]{regenauer2003review,regenauer2006effect,regenauer2008strain}. An accompanying criticism on the strain-weakening plasticity is that the softening rules often used in tectonic modeling lack reliable constraints. While agreeing that the lack of experimental and observational constraints on the strain-softening rules is problematic, we would like to point out that this problem does not invalidate the approach itself. In fact, shear band formation induced by strain softening is perfectly legitimate in the light of the 2nd law of thermodynamics. The source of previous confusion must be in the erroneous notion that only a strain hardening plasticity is ``stable'' in the sense that the plastic dissipation is positive therefore the 2nd law of thermodyanmics is satisfied. However, neither strain softening nor non-associated flow rule necessarily violates the maximum plastic dissipation postulate as shown by ~\citet[][see Sec. 3.2.]{Lubliner2008}

The maximum plastic dissipation postulate requires that the yield surface be convex in the principal stress space and the isotropic strain-softening plasticity model often used in tectonic modeling! satisfy this requirement by maintaining the convexity. For instance, when a strain-softening in the Mohr-Coulomb plastic model is realized through reduction in cohesion and friction angle, the hexagonal cone-shaped, therefore convex, yield surface in the principal stress space always remain convex except in the degenerate case where both friction angle and cohesion are zero. 

Under a typical setting for numerical tectonic modeling in which plastic flow is incompressible, i.e., dilation angle is 0$^{\circ}$, and the friction angle is about $30^{\circ}$, the principal stresses and plastic strain rates are indeed ``non-coaxial''~\citep{regenauer2003review,regenauer2006effect}. However, this fact does not necessarily mean that such a non-associated flow rule violates the maximum plastic dissipation. Friction angles usually assumed in tectonic models are never much greater than 30$^{\circ}$ and dilation angles are less than that. Since principal stress and plastic strain orientations are normal to a yield surface and a flow potential, respectively, the difference between their orientations is equal to that of the friction and the dilation angle. Because the double contraction of stress and plastic strain rate in the definition of plastic dissipation rate is equivalent to the projection operation, an acute angle between the principal stresses and plastic strain rates guarantees a positive plastic dissipation that indicates the satisfaction of the maximum plastic dissipation postulate. 

These considerations validate our adoption of the strain-weakening, non-associated, Mohr-Coulomb plastic model.

%********************** Benchmark *********
\section{Benchmarks}
\begin{figure}
   \label{fig:benchmark_model}
   \begin{center}
     \noindent \includegraphics[scale=1.0]{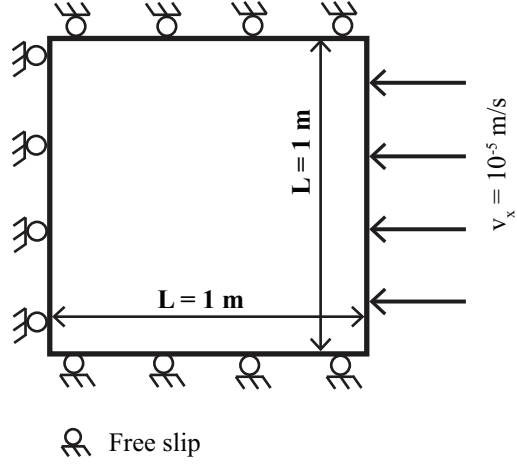}
   \end{center}
   \caption{Model geometry and boundary conditions for the plastic oedometer test. This figure and associated running/plotting scripts available under ~\citet{Ahamed2016}}
   \label{benchmarkDomain}
 \end{figure}

%------------- verify temperature solutions
%\subsection{Verification of temperature solutions}
We verify the governing equations implemented into \textit{DES3D}, an unstructured finite element code for long-term tectonic modeling~\citep{choi2013dynearthsol2d}, using three benchmark problems. The benchmarks are derived from the standard oedometer test~\citep[e.g.,][]{Davis2002, choi2013dynearthsol2d}. A cubic block of the Mohr-Coulomb plastic material is compressed in one direction while motions in the other directions are restricted (Fig. \ref{benchmarkDomain}). The symmetry and boundary conditions of the problem make it sufficient to discretize a square domain with two triangular elements. The simplicity of the problem allows for at least semi-analytic solutions. In benchmark-1, we include only the plastic power $(\boldsymbol{\sigma}:\dot{\boldsymbol{\epsilon}}_p)$ as a heat source and ignore the diffusion term. The density is assumed to be constant. Benchmark-2 solves the following equation
%
%--------------------- equation ------------------------------
\begin{equation} \label{bench_1}
(\rho c_p+ p\,\alpha_{v})\frac{dT}{dt}
=
\boldsymbol{\sigma}:\dot{\boldsymbol{\epsilon}}_p
+
T \alpha_{v} \frac{dp}{dt}
+
p\,T \alpha_{v} \nabla \cdot \boldsymbol{v},
\end{equation}
and the density is updated according to eq.~\eqref{24}.
Benchmark-3 verifies the thermal stress calculations in DES3D for a uniform temperature that increases linearly in time. We intentionally use a low density (1 kg/m$^{3}$) to make the contribution from plastic power non-negligible.
Parameters used in the benchmarks are listed in Table 1.

 % benchmark model setup

 % benchmark parameters
\begin{table}[h]
  \label{tab:parameters}
  \caption{Model parameters for Benchmarks}
  \centering
  \begin{tabular}{l c l}
    \hline
      Paramter & Symbol & Value \\
    \hline
      Bulk Modulus & $K$ & 200 MPa\\
      Shear Modulus & $G$ & 200 MPa\\
      Cohesion & $C$ & 1 MPa\\
      Friction Angle & $\phi$ & 10$^{\circ}$ \\
      Dilationa Angle & $\Psi$ & 10$^{\circ}$ \\
      Initail Temperature & $T_{0}$ & 273 K \\
      Reference density & $\rho_{0}$ & 1 kg/m$^{3}$ \\
      Volumetric expansion coefficient & $\alpha$ & 3.5 K$^{-1}$\\[1ex]
    \hline
  \end{tabular}
\end{table}

%---------------------------- benchmark-1 ----------------------------
\subsection{Analytic solutions for the oedometer test}
%The following analytical calculations are based on the oedometer setting depicted by figure \ref{fig:benchmark_model}.

\subsubsection{Total strain rate and strain}
The model geometry and boundary conditions give the following components of total strain rate ($\dot{\boldsymbol{\varepsilon}}$):
\begin{align}
  \dot{\varepsilon}_{xx} &= \frac{v_{x}}{L+v_{x}t}, \\
  \dot{\varepsilon}_{yy} &= 0, \\
  \dot{\varepsilon}_{zz} &= 0,
\end{align}
where $t$ is time, $v_x$ is the boundary velocity set to be $-10^{-5}$ m/s and $L$ is the edge length of the cube equal to 1 m.

The components of total strain ($\boldsymbol{\varepsilon}$) are given as
\begin{align}
  \varepsilon_{xx} &= \ln \left( \frac{L+v_{x}t}{L} \right), \\
  \varepsilon_{yy} &= 0, \\
  \varepsilon_{zz} &= 0.
\end{align}

\subsubsection{Pre-yielding stress}
Before yielding, stresses are updated according to the linear isotropic elasticity. In terms of the Lam\'e's constants ($\lambda$, $G$), stress components are given as:
\begin{align}
  \sigma_{xx} &= (\lambda+2G) \ln \left( \frac{L+v_{x}t}{L} \right), \\
  \sigma_{yy} &= \sigma_{zz} = \lambda \ln \left( \frac{L+v_{x}t}{L} \right).
\end{align}

\subsubsection{Mohr-Coulomb yield function and flow potential}
  We use the following forms of the Mohr-Coulomb yield function ($f$) and flow potential ($g$):
  \begin{align}
    f(\sigma_{xx},\sigma_{yy}) &= \sigma_{xx} - N_{\phi}\,\sigma_{yy} + 2C\,\sqrt{N_{\phi}}, \\
   g(\sigma_{xx},\sigma_{yy}) &= \sigma_{xx} - \frac{1 + \sin \psi}{1 - \sin \psi}\, \sigma_{yy}.
  \end{align}
where $N_{\phi}$ is defined as $(1 + \sin \phi)/(1 - \sin \phi)$, $\phi$ is the firction angle, $\psi$ is the dilation angle and $C$ is the cohesion.

\subsubsection{Yielding time ($t_{cr}$)} \label{t_cr}
We denote the time when yielding occurs for the first time as $t_{cr}$. Since $f(\sigma_{xx},\sigma_{yy}) = 0$ at $t=t_{cr}$,
\begin{equation}
  (\lambda+2G) \ln \left( \frac{L+v_{x}t_{cr}}{L} \right) - N_{\phi} \lambda \ln \left( \frac{L+v_{x}t_{cr}}{L} \right) + 2C\sqrt{N_{\phi}} = 0.
\end{equation}
Solving the above equation for $t_{cr}$, we get
\begin{equation}
  t_{cr} = \frac{L}{v_{x}} \left( \exp \left[ -\frac{2C\sqrt{N_{\phi}}}{(\lambda+2G) - N_{\phi}\lambda} \right] - 1 \right).
\end{equation}

\subsubsection{Plastic strain rate and strain}
We get the following expressions for plastic strain for $t\ge t_{cr}$:
\begin{align}
  {\varepsilon_{p}}_{xx} &= 2\beta(t), \\
  {\varepsilon_{p}}_{yy} &= -N_{\psi} \beta(t), \\
  {\varepsilon_{p}}_{zz} &= -N_{\psi} \beta(t),
\end{align}
where $\beta(t)$ is the plastic consistency paramter to be determined.
As in the classical plasticity theory~\citep[e.g.,][]{Lubliner2008}, the consistency parameter is zero before yielding ($t \le t_{cr}$).

The plastic strain rates are similarly defined in terms of the rate of consistency parameter ($\dot{\beta}(t)$):
\begin{align}
  \dot{{\varepsilon}_{p}}_{xx} &= 2\dot{\beta}(t), \\
  \dot{{\varepsilon}_{p}}_{yy} &= -N_{\psi} \dot{\beta}(t), \\
  \dot{{\varepsilon}_{p}}_{zz} &= -N_{\psi} \dot{\beta}(t).
\end{align}

\subsubsection{Elastic strain after yielding}
The above definitions of total and plastic strain lead to the following expressions for elastic strain as a function of time:
\begin{align}
  {\varepsilon_{e}}_{xx} &= \varepsilon_{xx} - {\varepsilon_{p}}_{xx} = \ln \left( \frac{L+v_{x}t}{L} \right) - 2\,\beta(t), \label{eps_el_xx} \\
  {\varepsilon_{e}}_{yy} &= \varepsilon_{yy} - {\varepsilon_{p}}_{yy} = N_{\psi} \, \beta(t), \label{eps_el_yy} \\
  {\varepsilon_{e}}_{zz} &= N_{\psi} \, \beta(t). \label{eps_el_zz}
\end{align}

\subsubsection{Post-yielding stress and determination of $\beta(t)$} \label{beta_section}

After yielding occurs, stresses are updated as follows:
\begin{align}
  \sigma_{xx} &= (\lambda+2G) \, {\varepsilon_{e}}_{xx} + \lambda \, ({\varepsilon_{e}}_{yy} + {\varepsilon_{e}}_{zz}),\\
  \sigma_{yy} &= (\lambda+2G) \, {\varepsilon_{e}}_{yy} + \lambda \, ({\varepsilon_{e}}_{zz} + {\varepsilon_{e}}_{xx}),\\
  \sigma_{zz} &= (\lambda+2G) \, {\varepsilon_{e}}_{zz} + \lambda \, ({\varepsilon_{e}}_{xx} + {\varepsilon_{e}}_{yy}).\\
\end{align}
Plugging \eqref{eps_el_xx} to \eqref{eps_el_zz} into the above equations, we get
\begin{align}
  \sigma_{xx}(t) &= (\lambda + 2G) \left( \ln \left( \frac{L+v_{x}t}{L} \right) - 2\beta(t) \right) + 2\lambda N_{\psi} \, \beta(t) \notag \\
&= (\lambda + 2G) \ln \left( \frac{L+v_{x}t}{L} \right) - \left[ 2(\lambda+2G) - 2\lambda N_{\psi} \right] \, \beta(t), \label{sxx_py} \\
  \sigma_{yy}(t) &= (\lambda + 2G) N_{\psi} \, \beta(t) + \lambda \left( N_{\psi} \, \beta(t) + \ln \left( \frac{L+v_{x}t}{L} \right) - 2\beta(t) \right) \notag \\
  &= \lambda \, \ln \left( \frac{L+v_{x}t}{L} \right) + \left[ 2(\lambda+G)N_{\psi}-2\lambda \right] \, \beta(t), \label{syy_py} \\
  \sigma_{zz}(t) &= \sigma_{yy}(t). \label{szz_py}
\end{align}

For $t \ge t_{cr}$, these stresses should always satisfy the yield condition:
\begin{equation}
  \sigma_{xx} - N_{\phi}\,\sigma_{yy} + 2C\,\sqrt{N_{\phi}} = 0.
\end{equation}
Substituting \eqref{sxx_py} and \eqref{syy_py} for the stress components in the yield condition, we get
\begin{equation}
\begin{split}
& (\lambda + 2G) \ln \left( \frac{L+v_{x}t}{L} \right) - \left[ 2(\lambda+2G) - 2\lambda N_{\psi} \right] \, \beta(t) \\
& - N_{\phi}\,\lambda \, \ln \left( \frac{L+v_{x}t}{L} \right) - \left[ 2(\lambda+G)\,N_{\phi}\,N_{\psi}-2\lambda\,N_{\phi} \right] \, \beta(t) + 2C\,\sqrt{N_{\phi}} = 0.
\end{split}
\end{equation}
Solving the above equation for $\beta(t)$, we get
\begin{equation}
\begin{split}
  \beta(t) &= \frac{ \left[ (\lambda + 2G) - N_{\phi} \lambda \right] \ln \left( \frac{L+v_{x}t}{L} \right) + 2C\,\sqrt{N_{\phi}}}{2(\lambda+G)N_{\phi}N_{\psi}+2(\lambda+2G)-2\lambda(N_{\phi}+N_{\psi})}.
\end{split}
\end{equation}
The time derivative of $\beta(t)$ is given as
\begin{equation}
\begin{split}
  \dot{\beta}(t) &= \frac{(\lambda + 2G) - N_{\phi} \lambda}{2(\lambda+G)N_{\phi}N_{\psi}+2(\lambda+2G)-2\lambda(N_{\phi}+N_{\psi})} \frac{v_{x}}{L+v_{x}t}.
\end{split}
\end{equation}

\subsubsection{Pressure and pressure rate}
Since $p = -(\sigma_{xx}+\sigma_{yy} +\sigma_{zz})/3$, we get the following from \eqref{sxx_py}, \eqref{syy_py} and \eqref{szz_py}:
\begin{align}
\begin{split}
  p(t) &= -K\,\ln \left( \frac{L+v_{x}t}{L} \right) + \left( 2\lambda+\frac{4}{3}G \right) (1-N_{\psi}) \, \beta(t) \\
  &= -K\,\ln \left( \frac{L+v_{x}t}{L} \right) + 2\,K\,(1-N_{\psi}) \, \beta(t), \\
  \dot{p}(t) &= - K \frac{v_{x}}{L+v_{x}t} + \left( 2\lambda+\frac{4}{3}G \right) (1-N_{\psi}) \, \dot{\beta}(t) \\
  &= - K \frac{v_{x}}{L+v_{x}t} + 2K \, (1-N_{\psi}) \, \dot{\beta}(t),
\end{split}
\end{align}
where $K$ is the bulk modulus defined as $\lambda + 2/3G$.

\subsubsection{Volume change rate}
\begin{equation}
  \nabla \cdot \mathbf{v} = \dot{\varepsilon}_{xx} + \dot{\varepsilon}_{yy} + \dot{\varepsilon}_{zz}= \frac{v_{x}}{L+v_{x}t}.
\end{equation}

\subsubsection{Density}

Using the expression for $\nabla \cdot \mathbf{v}$, the mass balance equation becomes
\begin{equation}
\begin{split}
  \frac{d\rho}{dt} &= -\rho \nabla \cdot \mathbf{v} = - \left( \frac{v_{x}}{L+v_{x}t} \right) \rho.
\end{split}
\end{equation}
%
%\begin{equation}
%  I = \exp \left[ \int \frac{v_{x}}{L+v_{x}t} \right] = e^{\ln (L+v_{x}t)} = L+v_{x}t.
%\end{equation}
%
%\begin{equation}
%  (I\,\rho)^{\prime} = 0
%\end{equation}
%
%\begin{equation}
%  I\,\rho = C_{1}
%\end{equation}
%
%\begin{equation}
%  \rho = \frac{C_{1}}{L+v_{x}t}
%\end{equation}
%
%Since $\rho(0) = \rho_{0}$,
%
%\begin{equation}
%  \rho_{0} = \frac{C_{1}}{L}.
%\end{equation}
%
%\begin{equation}
%  C_{1} = \rho_{0}\,L.
%\end{equation}
%
Solving for $\rho(t)$, we get
\begin{equation}
  \rho(t) = \rho_{0} \frac{L}{L+v_{x}t},
\end{equation}
where $\rho_{0}$ is the reference density at $t=0$.
%We use $\rho_{0}=1$ kg/m$^{3}$ so that the coefficient $\rho\,c_{p}$ does not make the heat source terms negligible.

\subsection{Benchmark-1 : Plastic power only, no thermal stress}
The equation we are going to solve is
\begin{equation}
  \rho c_{p} \frac{dT}{dt} = \boldsymbol{\sigma}:\dot{\boldsymbol{\varepsilon}}_{p},
\end{equation}
where
\begin{equation}
\begin{split}
  \boldsymbol{\sigma}:\dot{\boldsymbol{\varepsilon}}_{p} &= \sigma_{xx} \, \dot{{\varepsilon}_{p}}_{xx} + \sigma_{yy} \, \dot{{\varepsilon}_{p}}_{yy} + \sigma_{zz} \, \dot{{\varepsilon}_{p}}_{zz} \\
 &= 2 \, \dot{\beta}(t) \, \sigma_{xx}(t) - 2 \, N_{\psi} \, \dot{\beta}(t) \, \sigma_{yy}(t) \\
 &= 2 \, \dot{\beta}(t) \, \left[ \sigma_{xx}(t) - N_{\psi} \, \sigma_{yy}(t) \right].
\end{split}
\end{equation}
Since $\rho$ and $c_{p}$ are assumed to be constant,
\begin{equation}
\begin{split}
  \frac{dT}{dt} &= \frac{1}{\rho c_{p}} \boldsymbol{\sigma}:\dot{\boldsymbol{\varepsilon}}_{p} \\
                &= \frac{2\dot{\beta}(t)}{\rho c_{p}} \left[ \sigma_{xx}(t) - N_{\psi}\,\sigma_{yy}(t) \right].
\end{split}
\end{equation}
We use the forward Euler scheme to integrate the above time derivative of $T$ with the initial condition $T(0)=273$ K.

\begin{figure} \label{f1}
 \begin{center}
 \noindent \includegraphics[width=0.8\textwidth]{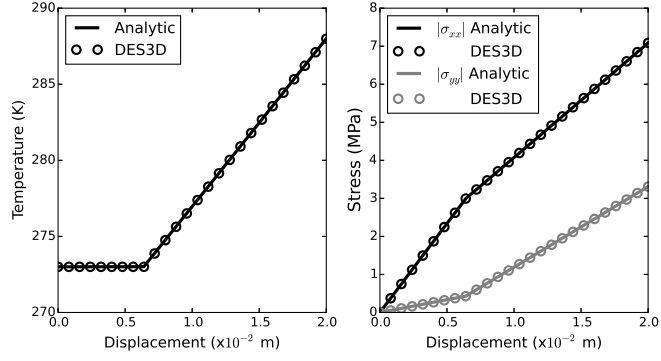}
 \end{center}
 \caption{Temperature ($T$), Stress xx ($|\boldsymbol{\sigma}_{xx}|$), and Stress yy ($|\boldsymbol{\sigma}_{yy}|$) are plotted against x-displacement from analytic and numerical solutions by \textit{DES3D} for benchmark-1. This figure and associated running/plotting scripts available under ~\citet{Ahamed2016}}
 \label{benchmark-1}
\end{figure}

Results of the test are shown in Fig. \ref{benchmark-1}. The relative error of the DES3D solution relative to the semi-analytic one is 0.004 \%.
%
% Benchmark-1

%---------------------------- benchmark-2 ----------------------------
\subsubsection{Benchmark-2 : The full equation, no thermal stress}
The equation to solve is

\begin{equation}
 (\rho\,c_{p}+p\,\alpha_{v})\frac{dT}{dt} = \boldsymbol{\sigma}:\dot{\boldsymbol{\varepsilon}}_{p} + T\,\alpha_{v}\,\frac{dp}{dt} + p\,T\,\alpha_{v}\,\nabla \cdot \mathbf{v}.
\end{equation}
Using the derived expressions for the involved quantites and the forward Euler scheme, we can integrate the above equation for the initial condition, $T(0)=273.0$ K.

\begin{figure} \label{f2}
 \begin{center}
 \noindent \includegraphics[width=0.8\textwidth]{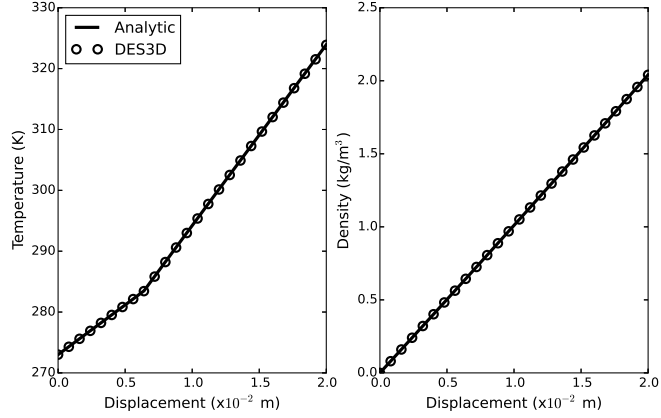}
 \end{center}
 \caption{Temperature ($T$) and Density ($\rho$) are plotted against x-displacement from analytic and numerical solutions by \textit{DES3D} for benchmark-2. This figure and associated running/plotting scripts available under ~\citet{Ahamed2016}}
 \label{benchmark-2}
\end{figure}

Results of the test are shown in Fig. \ref{benchmark-2}. The relative errors of the temperature and density from DES3D are 0.01 \% and 0.00003 \%, respectively.

% Benchmark-2

%---------------------------- benchmark-3 ----------------------------
\subsection{Benchmark-3: Thermal stresses under a prescribed temperature change}

We verify the implementation of the thermoelastic constitutive equations in DES3D.
For simplicity, we prescribe a uniform temperature field, which is initially 273 K and increases linearly in time as
\begin{align} \label{b3-90}
T(t) = b\,t,
%\Delta T &= 2\times 10^{-2}\Delta t, \notag \\
%T^{t+\Delta t} &=T^t+2\times 10^{-2}\Delta t.
\end{align}
where $b$ is a constant. We set $b$ to be 0.4 K/s.

Under this setting, we can analytically derive the elastic and plastic stress solutions for the oedometer test involving thermal stress.
In the elastic regime, the stresses are given as
\begin{align}
  \sigma_{xx} &= (\lambda+2G) \, {\varepsilon_{e}}_{xx} + \lambda \, ({\varepsilon_{e}}_{yy} + {\varepsilon_{e}}_{zz}) - K \alpha_{v} b\,t,\\
  \sigma_{yy} &= (\lambda+2G) \, {\varepsilon_{e}}_{yy} + \lambda \, ({\varepsilon_{e}}_{zz} + {\varepsilon_{e}}_{xx}) - K \alpha_{v} b\,t,\\
  \sigma_{zz} &= (\lambda+2G) \, {\varepsilon_{e}}_{zz} + \lambda \, ({\varepsilon_{e}}_{xx} + {\varepsilon_{e}}_{yy}) - K \alpha_{v} b\,t.
\end{align}
Using the expressions, \eqref{sxx_py}, \eqref{syy_py} and \eqref{szz_py}, we get
\begin{align}
  \sigma_{xx}(t) &= (\lambda + 2G) \left( \ln \left( \frac{L+v_{x}t}{L} \right) - 2\beta(t) \right) + 2\lambda N_{\psi} \, \beta(t) - K \alpha_{v} b\,t \\
&= (\lambda + 2G) \ln \left( \frac{L+v_{x}t}{L} \right) - \left[ 2(\lambda+2G) - 2\lambda N_{\psi} \right] \, \beta(t) - K \alpha_{v} b\,t, \\
  \sigma_{yy}(t) &= (\lambda + 2G) N_{\psi} \, \beta(t) + \lambda \left( N_{\psi} \, \beta(t) + \ln \left( \frac{L+v_{x}t}{L} \right) - 2\beta(t) \right) - K \alpha_{v} b\,t \\
  &= \lambda \, \ln \left( \frac{L+v_{x}t}{L} \right) + \left[ 2(\lambda+G)N_{\psi}-2\lambda \right] \, \beta(t) - K \alpha_{v} b\,t, \\
  \sigma_{zz}(t) &= \sigma_{yy}(t).
\end{align}
We follow the procedure in Sec.~\ref{t_cr} to compute the timing of the first yielding, $t_{cr}$, but use the Newton method
because the equation for $t_{cr}$ does not allow a solution in terms of elementary functions.
Following the procedure for determining $\beta(t)$ and $\dot{\beta}(t)$ described in Sec. \ref{beta_section}, we get
%%
%\begin{equation}
%\begin{split}
%&(\lambda + 2G) \ln \left( \frac{L+v_{x}t}{L} \right) - K \alpha_{v} b\,t - \left[ 2(\lambda+2G) - 2\lambda N_{\psi} \right] \, \beta(t) \\
%-&N_{\phi}\,\lambda \, \ln \left( \frac{L+v_{x}t}{L} \right) + N_{\phi}\,K \alpha_{v} b\,t - \left[ 2(\lambda+G)\,N_{\phi}\,N_{\psi}-2\lambda\,N_{\phi} \right] \, \beta(t) + 2C\,\sqrt{N_{\phi}} = 0.
%\end{split}
%\end{equation}
%
\begin{equation}
\begin{split}
  \beta(t) &= \frac{ \left[ (\lambda + 2G) - N_{\phi} \lambda \right] \ln \left( \frac{L+v_{x}t}{L} \right) + (N_{\phi}-1)\,K \alpha_{v}\,b\,t + 2C\,\sqrt{N_{\phi}}}{2(\lambda+G)N_{\phi}N_{\psi}+2(\lambda+2G)-2\lambda(N_{\phi}+N_{\psi})},
\end{split}
\end{equation}
and
\begin{equation}
\begin{split}
  \dot{\beta}(t) &= \frac{ \left[ (\lambda + 2G) - N_{\phi} \lambda \right] \frac{v_{x}}{L+v_{x}t} + (N_{\phi}-1)\,K \alpha_{v}\,b}{2(\lambda+G)N_{\phi}N_{\psi}+2(\lambda+2G)-2\lambda(N_{\phi}+N_{\psi})} .
\end{split}
\end{equation}
As before, $\beta(t) = \dot{\beta}(t) = 0$ for $t < t_{cr}$.

\begin{figure} \label{f3}
 \begin{center}
 \noindent \includegraphics[width=0.75\textwidth]{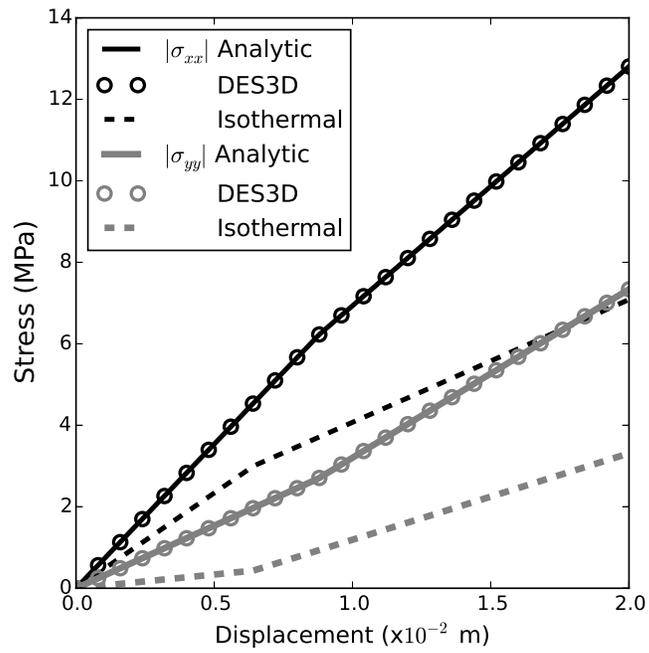}
 \end{center}
 \caption{$|\boldsymbol{\sigma}_{xx}|$ and $|\boldsymbol{\sigma}_{yy}|$ from the analytic (circles) and the DES3D numerical solution (solid lines) for benchmark-3,  plotted against displacement. The corresponding analytic solutions for the isothermal case (dashed lines) are shown for comparison. This figure and associated running/plotting scripts available under ~\citet{Ahamed2016}}
 \label{benchmark-3}
\end{figure}

Differential stress arising in this benchmark is the same with those of the isothermal case but pressure in this test is greater due to the compressional thermal stress caused by the prescribed temperature increase. As a result, the first yielding in benchmark-3 should occur at a greater value of differential stress, or equivalently, displacement than in the isothermal case. In other words,  than in the isothermal case. The results of benchmark-3 shows in Fig. \ref{benchmark-3} are consistent with this expectation. The relative errors of $\sigma_{xx}$ and $\sigma_{yy}$ computed with DES3D are 0.0028 \% and 0.01 \%.

% Benchmark-3

%---------------------------- Results and discussion ----------------------------
\section{Discussion}
Using the verified implementation of the ``full'' energy balance equation,~\eqref{25}, as well as the thermo-elasticity, we now systematically assess the impact of the thermo-mechanically coupled governing equations on a more geologically relevant problem. In this study, we choose the large-offset normal fault evolution~\citep[e.g.][]{lavier2000factors} as an example. Studied extensively and understood well, this system will facilitate identification and attribution of differences between the models of the newly-introduced physics and the uncoupled, isothermal ones.

\begin{figure}[h!]
\begin{center}
\includegraphics[width=1.0\textwidth]{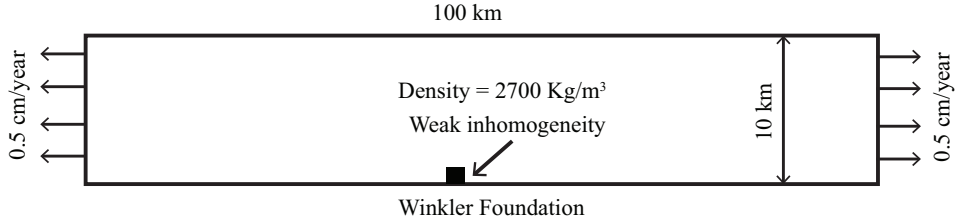}
\end{center}
\caption{Setup of the large offset normal fault models. This figure and associated running/plotting scripts available under ~\citet{Ahamed2016}}
\label{fig:faultModelSetup}
\end{figure}

%\subsection{Description of the normal fault models}
%\subsection{Normal fault model setup}
We create a normal fault in an extending Mohr-Coulomb elasto-plastic layer which is initially 100 km long and 10 km thick (Fig.~\ref{fig:faultModelSetup}). Both sides of the layer are pulled at a constant velocity of 0.5 cm/yr. The bottom boundary is supported by the Winkler foundation \citep[pp.95]{watts2001isostasy}. To induce strain localization, we decrease cohesion from 40 MPa to 8 MPa linearly as plastic strain increases to 1.
% to perscribed threshold value\citep[e.g]{lavier2000factors}.
We add a weak zone at the bottom center of the domain to initiate a fault from there.
%To prevent numerical difficulty involved with meshing a highly rugged topography,
We impose topographic smoothing of the diffusion type with a transport coefficient of $10^{-7} m^2/s$ \citep[pp. 225]{turcotte2014geodynamics}.
%Figure \ref{fig:faultModelSetup} shows the model set up and
%Table-\ref{tab:parameters2} shows the list of the parameters used to run these models.

%% ------- figure : faultModelSetup ----------------

We set up five models that differ in the form of the energy balance equation and in the presence of volumetric plastic strain.
In model-1, we do not consider any heat-generating mechanism. As a result, the initial temperature, which is uniformly $0 ^{\circ}$C, does not change in time. The behavior of model-1 should be similar to that of the ``unlimited, footwall-snapping'' mode of~\citep{lavier2000factors}.
%
%\begin{equation}
%  \rho \, c_p \, \frac{dT}{dt} = -\nabla \cdot \boldsymbol{q}.
%\end{equation}
%
In the rest of models (model-2 to 5) we solve the full energy balance equation~\eqref{25} with the heat source/sink term, $s$, equal to zero and with the temperature feedback to rheology through thermo-elasticity. To explore the effects of volumetric plastic strain, we use a non-zero dilation angle in model-3 to 5. The dilation angle is fixed in model-3 while we reduce it to zero as the accumulated plastic strain increases to a prescribed value of 1.0 in model-4 and 5. The model-5 is the same as model-4 but we intentionally include only the deviatoric part of the plastic power ($\boldsymbol{\sigma}:\dot{\boldsymbol{\varepsilon}}_{p}$) in equation \eqref{25}. By comparing model-4 and 5, we can decide whether the volumetric plastic power can be ignored. Table 2 summarizes the differences among the models. Table 3 shows the list of the parameters used in the models.
%
%  model descriptions
\begin{table}[h!]
  \label{tab:model_descriptions}
  \caption{Description of the normal fault evolution models }
  \centering
  \begin{tabular*}{\textwidth}{@{\extracolsep{\fill}} l c c c c }
    \hline
    Models
    & Energy balance
    & Dilation angle
    & Plastic power\\
    \hline
      Model-1 & Heat diffusion only & 0$^\circ$ & N/A\\
      Model-2 & Full & 0$^\circ$ & Total \\
      Model-3 & Full & 10$^\circ$ & Total \\
      Model-4 & Full & 10$^\circ$ to 0$^\circ$ & Total \\
      Model-5 & Full & 10$^\circ$ to 0$^\circ$ & Deviatoric \\[1ex]
    \hline
  \end{tabular*}
\end{table}

%
%
%  model parameters
\begin{table}[h]
  \label{tab:fault_parameters}
  \caption{Parameters for the normal fault evolution models}
  \centering
  \begin{tabular}{l c c}
  \hline
  Parameter & Symbol & Value \\
  \hline
  Bulk Modulus & $K$ & 50 GPa\\
  Lam\'{e}'s constant & $\lambda$ & 30 GPa \\
  Shear Modulus & $G$ & 30 GPa\\
  Initial Cohesion & $C$ & 40 MPa\\
  Friction Angle & $\phi$ & $30^{\circ}$ \\
  Dilation Angle & $\Psi$ & $10^{\circ}$ \\
  Density & $\rho$ & 2700 $Kg/m^3$ \\
  Volumetric expansion coefficient & $\alpha$ & 3.5 $K^{-1}$ \\[1ex]
  \hline
  \end{tabular}
\end{table}

We see noticeable differences in fault geometry and shape of the core complex between model-1 and model-2 that solves the full energy balance equation and includes thermal stresses.
%Figures \ref{fig:diff_and_energy} to \ref{fig:fixed_reduced_dev} show the plastic strain distribution of all the models.
The overall behavior of the faults in model-1 is similar to the unlimited, footwall snapping mode of~\citet{lavier2000factors}.
The geometry of the primary fault of model-2 is almost the same as that of the model-1 until about 17 km of extension.
Around this time, however, model-2 forms a secondary fault but model-1 does not yet.
%The primary fault of model-1 and 2 become significantly different between 20 and 40 km extension.
%The geometry of the secondary fault started to differ since their formation.
After 20 km of extension, more secondary faults start form in both models but their geometry and location are not identical.
For instance, a fault block forms in model-2 after about 30 km of extension when a new high-angle fault forms next to the first secondary fault.
Model-1 does not develop a corresponding structure including a fault block after the same amount of extension.
Model-1 forms a single secondary fault initially, which eventually connects with the second secondary fault after about 40 km of extension.
At 30 km of extension model-2 has three well-developed secondary faults while model-1 has two.
Initially, the shape of core complex remains same until 10 km of extension. Because of different fault geometry the shape of the core complex started to differ from each other from 20 km of extension. At 40 km of extension the core-complex of the model-2 becomes elongated with the almost flat surface while the model-1 has more rounded one.

\begin{figure}[h!]
\begin{center}
\includegraphics[width=1.0\textwidth]{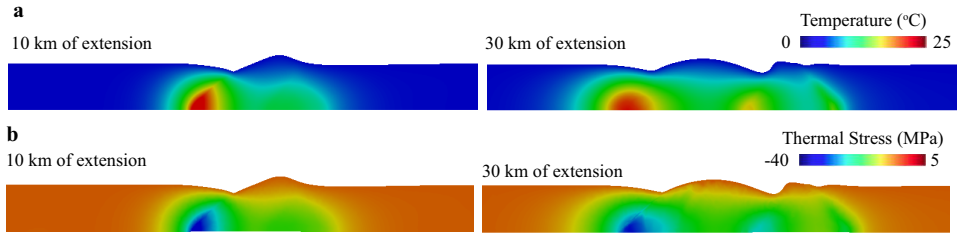}
\end{center}
\caption{(a) Thermal stress and (b) temperature distribution of model-2 after 10 and 30 km of extension. This figure and associated running/plotting scripts available under ~\citet{Ahamed2016}}
\label{fig:stress_temp}
\end{figure}

Figure \ref{fig:stress_temp} shows the distribution of temperature in model-2 and the corresponding thermal stresses after 10 and 30 km of extension. The temperature increase near the active faults are  greater than 15 $^{\circ}$C and the magnitude of thermal stress is as large as 40 MPa. The thermal stress magnitude is a non-negligible fraction, about 10 \%, of lithostatic stress near the bottom. Thus, the different behaviors of the models can be attributed to the presence of thermal stresses only in model-2.
%% ------- figure : temperatureDistribution -----------
\begin{figure}[h!]
\begin{center}
\includegraphics[width=1.0\textwidth]{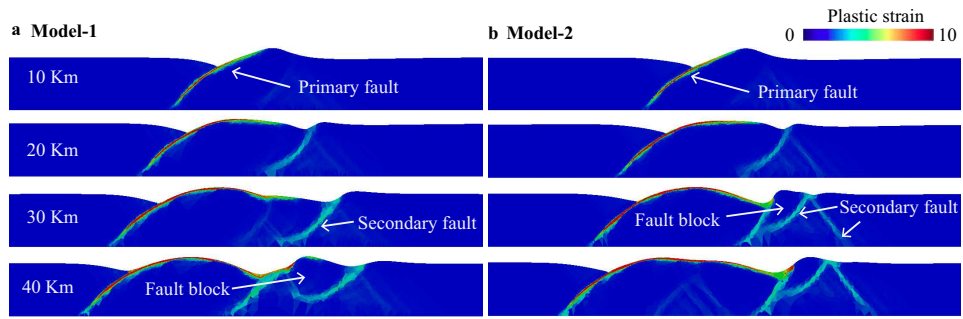}
\end{center}
\caption{Plastic strain distribution of (a) model-1 and (b) model-2 after 10, 20, 30 and 40 km of extensions. This figure and associated running/plotting scripts available under ~\citet{Ahamed2016}}
\label{fig:diff_and_energy}
\end{figure}

%\subsection{Effects of fixed dilation angle}
%All of the models with a non-zero dilation angle prefer forming a rounded and elevated single core-complex.
Identical with model-2 except a non-zero dilation angle of 10$^{\circ}$, model-3 develops an unrealistically elevated core-complex with a relief greater than 10 km and shear bands that are much wider than those in model-1 and 2. Expansion of shear bands when dilation angle is non-zero is kinematically expected. The dilation angle fixed at 10$^{\circ}$ sustains a higher level of plastic power than the non-dilational cases zero dilation angle. The greater plastic power leads to a greater amount of temperature change and thermal stresses. The thermal stresses pushes up the free-traction top surface, creating the highly elevated core-complex.
%The overly expanded shear bands of this model are not consistent with the shear bands found in nature.
%Therefore, keeping the dilation angle constant might produce unrealistic results.

%\subsection{Effects of volumetric plastic power}
%The volumetric plastic power with deviatoric power can change fault geometry, and tectonic evolution significantly.

%% ------- figure : plasticStrain -------------------
\begin{figure}[h!]
\begin{center}
\includegraphics[width=0.8\textwidth]{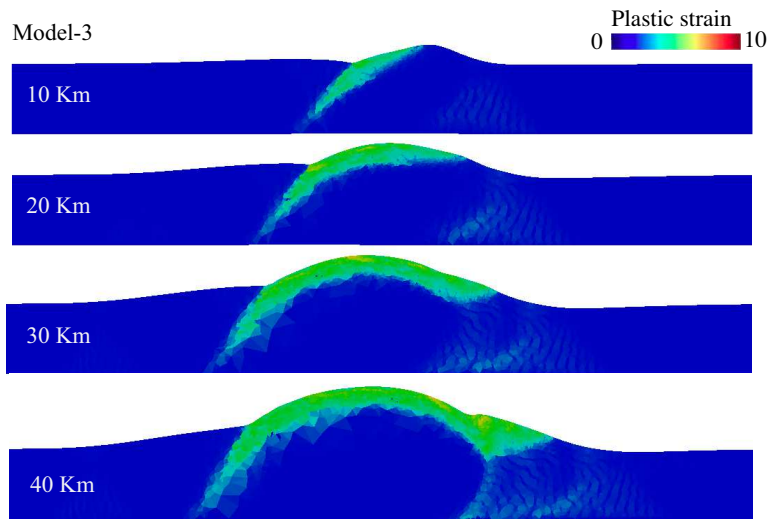}
\end{center}
\caption{Same with Fig.~\ref{fig:diff_and_energy} but for model-3. This figure and associated running/plotting scripts available under ~\citet{Ahamed2016}}
\label{fig:fixed}
\end{figure}

\begin{figure}[h!]
\begin{center}
\includegraphics[width=1.0\textwidth]{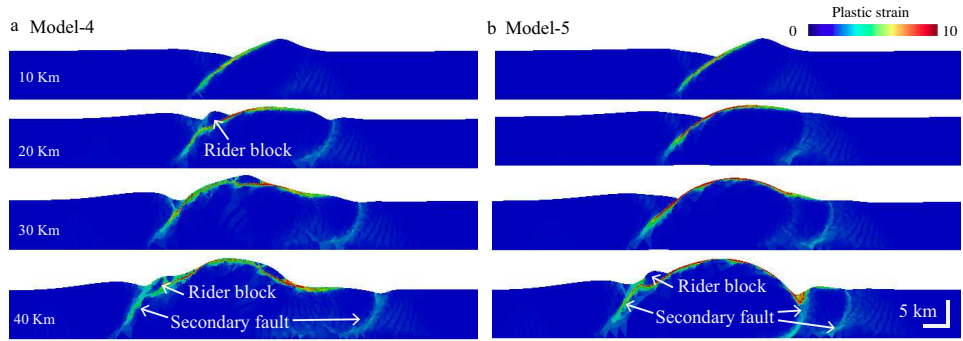}
\end{center}
\caption{Same with Fig.~\ref{fig:diff_and_energy} but for (a) model-4 and (b) model-5. This figure and associated running/plotting scripts available under ~\citet{Ahamed2016}}
\label{fig:dev_and_reduce}
\end{figure}

To isolate the effect of volumetric plastic power, we construct the model-4 and 5 that are identical except that model-4 considers the total plastic power while model-5 includes only the deviatoric part of the plastic power. In order to prevent the unrealistic expansion of shear bands as in model-3, the dilation angle is reduced with accumulated plastic strain in both model-4 and 5~\citep{choi2015making}.

Models-4 and 5 do not show notable differences until 10 km extension  (Figure \ref{fig:dev_and_reduce}).  However, after about 20 km of extension, they start to exhibit differences in the fault geometry as well as in the timing of rider block formation. A rider block starts to form at 17 km of extension in model-4 but model-5 forms one only after more than 37 km of extension. After 40 km of extension, model-5 has three secondary faults produced by the footwall snapping but model-4 still has only two secondary fault. The topographic relief in model-4 and 5 is about 5 km, which is much greater than the relief of the previous models for the large offset normal fault~\citep{lavier2000factors,choi2013a} but about half of that of model-3 with a fixed dilation angle of 10$^{\circ}$. The excessive reliefs in the models with non-zero dilation angle, model-3 to 5, suggest that the rate of dilation angle reduction should be greater than that of this study to have 1-2 km topographic relief as in the previous studies on large-offset normal faults. The noticeable differences between model-4 and 5 suggest to include the total plastic power in the energy balance rather than only the deviatoric component.
Ignoring the volumetric plastic power as in model-5 is inconsistent with the kinematics in the first place.

%

%%%% Figures for this section %%
%%% ------- figure : faultModelSetup ----------------
%\begin{figure}[h!]
%\begin{center}
%\includegraphics[width=15cm]{figures/faultModelSetup.png}
%\end{center}
%\caption{Model setup for large offset normal fault}
%\label{fig:faultModelSetup}
%\end{figure}
%%
%%% ------- figure : plasticStrain -------------------
%\begin{figure}[h!]
%\begin{center}
%\includegraphics[width=15cm]{figures/fixed_reduced_dev.png}
%\end{center}
%\caption{Non-zero dilation angle models (a) model-3 and (b) model-4 and (c) model-5}
%\label{fig:fixed_reduced_dev}
%\end{figure}
%%
%%% ------- figure : temperatureDistribution -----------
%\begin{figure}[h!]
%\begin{center}
%\includegraphics[width=15cm]{figures/diff_and_energy.png}
%\end{center}
%\caption{Zero dilation angle models (a) model-1 and (b) model-2}
%\label{fig:diff_and_energy}
%\end{figure}

\section{Conclusions}
We derive a temperature evolution equation based on the energy balance principle that accounts for deformation-related energy changes as well as the conventional heat energy diffusion and advection. An explicit-time finite element solution procedure for the derived equation is implemented in DES3D, an unstructured mesh finite element solver for geodynamics. We verify the implementation using three benchmark tests that have semi-analytic solutions. Benchmark-1 includes only the total plastic power term in the energy balance equation. Benchmark-2 includes the full energy equation as well as the mass balance equation. Benchmark-3 prescribes a temperature change linear in time and computes thermal stresses. The numerical solutions from DES3D for all the benchmarks show an excellent match with the corresponding semi-analytic solutions. Coupling the verified implementation of the full temperature evolution with the strain weakening Mohr-Coulomb plastic rheology through thermal stresses, we also explore the long-term behaviors of a large-offset normal fault. We find that the temperature arising mostly from the inelastic power term in the full energy balance has non-negligible effects on the long-term evolution of normal fault. For instance, when the plastic power is considered, temperature increases by more than 15$^{\circ}$C and the magnitude of the associated thermal stress is as large as 40 MPa. These extra stresses appear to promote the formation of secondary faults and an elongated core complex. When the dilational plastic deformation is enabled, our models develop even more differences in faulting behaviors such as the formation of a rider block and the great topographic relief, compared to the previous models for the large-offset normal fault with the same parameters and geometry. We conclude that the new coupled governing equations presented in this study has significant impact on long-term tectonics. Although not so strong as to cause a fundamental shift in faulting modes under tested conditions, the effects of deformation energetics might be critically important in other geologically relevant systems and conditions.

\section*{References}

\bibliography{refs}

%\section{APENDIX}

\end{document}